%% file: EASE19_main.tex
\documentclass[sigconf]{acmart}

\setcopyright{rightsretained}

\acmDOI{10.475/123_4}

\acmISBN{123-4567-24-567/08/06}

\begin{document}
\title{SZZ Unleashed: An Open Implementation of the SZZ Algorithm}
\subtitle{-- Featuring Example Usage in a Study of Just-in-Time Bug Prediction for the Jenkins Project}

\author{Markus Borg}
\affiliation{%
  \institution{ICT SICS\\RISE Research Institutes of Sweden AB}
  \city{Lund}
  \country{Sweden}
}
\email{markus.borg@ri.se}

\author{Oscar Svensson}
\affiliation{%
  \institution{Dept. of Computer Science\\Lund University}
  \city{Lund}
  \country{Sweden}
}
\email{wgcp92@gmail.com}

\author{Kristian Berg}
\affiliation{%
  \institution{Dept. of Computer Science\\Lund University}
  \city{Lund}
  \country{Sweden}
}
\email{kristianberg.jobb@gmail.com}

\author{Daniel Hansson}
\affiliation{%
  \institution{Verifyter AB}
  \city{Lund}
  \country{Sweden}
}
\email{daniel.hansson@verifyter.com}

\renewcommand{\shortauthors}{M. Borg \textit{et al.}}

\begin{abstract}
Numerous empirical software engineering studies rely on detailed information about bugs. While issue trackers often contain information about when bugs were fixed, details about when they were introduced to the system are often absent. As a remedy, researchers often rely on the SZZ algorithm as a heuristic approach to identify bug-introducing software changes. Unfortunately, as reported in a recent systematic literature review, few researchers have made their SZZ implementations publicly available. Consequently, there is a risk that research effort is wasted as new projects based on SZZ output need to initially reimplement the approach. Furthermore, there is a risk that newly developed (closed source) SZZ implementations have not been properly tested, thus conducting research based on their output might introduce threats to validity. We present SZZ Unleashed, an open implementation of the SZZ algorithm for git repositories. This paper describes our implementation along with a usage example for the Jenkins project, and conclude with an illustrative study on just-in-time bug prediction. We hope to continue evolving SZZ Unleashed on GitHub, and warmly invite the community to contribute. 
\end{abstract}

%
%



\maketitle

\input{EASE19_body.tex}

\bibliographystyle{ACM-Reference-Format}
\bibliography{szz}

\end{document}

%% file: EASE19_body.tex



\section{Introduction}
Empirical software engineering research often rely on detailed bug information. Bug information is often maintained in issue trackers such as Jira or BugZilla, which has enabled numerous publications related to mining software repositories~\cite{cavalcanti_challenges_2014,de_freitas_farias_systematic_2016}. However, while issue trackers often contain details about both bugs (e.g., version information, references to failed test case executions) and the subsequent bug fixes (e.g., who developed the fix and a reference to a specific commit with the resolution), information about the root cause of a bug and when it was introduced are often missing. 

In many software engineering research studies, knowing which individual commit that introduced a bug is essential -- examples include work on fault prediction~\cite{hall_systematic_2012}, test case selection~\cite{engstrom_systematic_2010}, and static code analysis~\cite{rahman_comparing_2014}. One approach to address missing bug information is to heuristically deduce it. Successful approaches to extend the information stored in issue trackers can be of great value to empirical software engineering. However, for such an approach to be useful, it has to deliver reliable output that both industry and academia trust~\cite{fenton_critique_1999,czerwonka_crane:_2011}.

A popular approach to extend bug information is to propose ``bug-introducing changes'' for the existing Bug Reports (BR). The dominant algorithm to do this is called SZZ, after the three authors of the seminal paper~\cite{sliwerski_when_2005}: {\'S}liwerski, Zimmermann, and Zeller. In a recent study on reproducability and credibility of software engineering research, Rodriguez-Perez \textit{et al.} presented a systematic literature review on research that used SZZ~\cite{rodriguez-perez_reproducibility_2018}. They identified 187 studies, and found that researchers typically implement their own versions of SZZ rather than building on what others have previously done. Rodriguez-Perez \textit{et al.} suggest that one reason is that researchers rarely make the SZZ implementations publicly available, thus any researcher relying on SZZ must first implement it from scratch. While there are some partial SZZ implementations available~\cite{rosen_commit_2015,correia_old-szz_2017}, Rodriguez-Perez \textit{et al.} call for researchers to publish source code to allow others to fork the project.

In this paper, we respond to Rodriguez-Perez \textit{et al.}'s call for improved reproducability through an open source implementation of SZZ. We introduce \textit{SZZ Unleashed} -- available on GitHub under an MIT license since June 2018~\cite{svensson_szz_2018}. The source code was developed as part of a MSc. thesis project at Axis Communications AB in Lund, Sweden. We have tested SZZ Unleashed on the repository of the Jenkins automation server\footnote{https://github.com/jenkinsci/jenkins} and used the results to train a random forest classifier for Just-In-Time (JIT) bug prediction~\cite{kamei_large-scale_2013}, i.e., to identify high-risk changes at commit-time. At the time of this writing, SZZ Unleashed has been forked four times -- at least twice by senior researchers from academia -- and we have approved the first external pull request. Since members of the research community have already found and forked SZZ Unleashed, we conclude that there is a demand for our implementation.

The rest of the paper is structured as follows. In Section~\ref{sec:bg} we introduce the SZZ algorithm and some later improvements. Section~\ref{sec:impl} presents the implementation of SZZ Unleashed along with an example for the Jenkins project. In Section~\ref{sec:example}, we illustrate how the SZZ Unleashed output for the Jenkins project can be used, by training a random forest classifier for JIT bug prediction. Finally, Section~\ref{sec:conc} concludes the paper and presents how we would like SZZ Unleashed to evolve.

\section{Background -- The SZZ algorithm}\label{sec:bg}
The SZZ algorithm was developed as an approach to identify bug-introducing commits in a software repository. It was introduced by {\'S}liwerski \textit{et al.} \cite{sliwerski_when_2005}, and was later given its name after the initials of the three authors. While the SZZ algorithm was developed for the CVS version control system and its corresponding commit practices, SZZ has evolved also for software repositories that use git. The SZZ algorithm is organized in two subsequent phases.

In the first phase, BRs in the issue tracker are linked to bug-fixing commits. This is done by using regular expressions to find explicit references to BRs in commit messages. If the content of the issue tracker is less structured, then commit messages that contain the word ``fix'' -- or whatever convention is used in the project under study -- are assumed to be bug fixes. For each of the bug-fixing commits that were identified, the modified lines in the source code are extracted.

\begin{figure*}
\centering
\includegraphics[width=\textwidth]{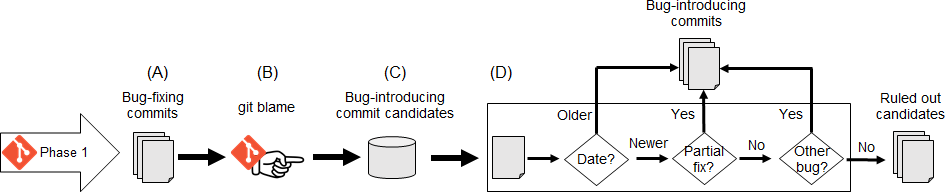}
\caption{Overview of the second phase of the SZZ algorithm.}
\label{fig:szz}
\end{figure*}

Figure~\ref{fig:szz} shows the steps in the second phase. For each bug-fixing commit from the first phase (A), SZZ uses the \textit{git blame} command (B) to identify all commits that previously made changes to the same lines of code. Git blame shows what revision and author last modified each line of a file, i.e., executing git blame on a bug-fixing commit results in a set of commits that might have introduced the bug. We refer to these a bug-introducing commit candidates (C).

For each candidate, SZZ determines whether it can be ruled out as bug-introducing or not (D). First, the commit time of a candidate is compared to the time when the corresponding BR was submitted. If the commit time is later than the report submission time, the candidate can be bug-introducing only if it is 1) a partial fix, i.e., a fix that did not completely resolve the bug it intended to resolve -- as made evident by a later bug-fixing commit for the same issue, or if it is 2) responsible for another bug, i.e., the candidate is responsible for a bug different from the one resolved by the bug-fixing commit that blamed the candidate. This means that another bug-fixing commit can have its bug origin in this commit because they have both made changes to the same file. We present more details in Section~\ref{sec:impl}, where we describe our implementation of SZZ.

\begin{figure}
\centering
\includegraphics[width=0.75\columnwidth]{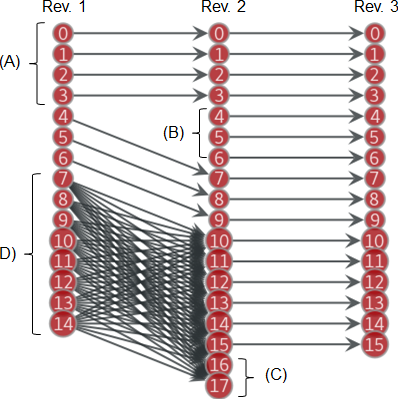}
\caption{Annotation graph mapping lines of source code between revisions.}
\label{fig:anno}
\end{figure}

Kim \textit{et al.} presented improvements to the SZZ algorithm~\cite{kim_automatic_2006}, including \textit{annotation graphs} created by origin analysis~\cite{godfrey_using_2005} and an approach to filter out cosmetic changes to source code. Figure~\ref{fig:anno} presents an annotation graph for a source code file, created by mapping different revision of the source code by using the \textit{git annotate} command (exists also in CVS and SVN, now replaced by \textit{git blame}). Each node shows a version of a single line of code and edges illustrate relations between revisions. The first four lines (A) are unmodified between the three revisions. Nodes 4--6 in revision 2 did not exist in the revision 1, i.e., these lines of code were inserted. Consequently, the fifth line in revision 1 (node 4) is instead mapped to node 7 in revision 2. Nodes 16 and 17 in Revision 2 are not mapped to any lines in Revision 3, i.e., they were deleted. 

If multiple adjacent lines are changed, then the annotation graph technique will fail to map those lines between revisions. Adjacent lines of code that are modified map will simply be mapped to the same set of lines in the next revision. This can be seen for nodes 7--14 in Revision 1, which are mapped to nodes 10--17 in revision 2. Consequently, SZZ relying on annotation graphs is not very precise in tracing changes across revisions.

Williams and Spacco addressed the lack of SZZ precision by replacing the annotation graph with an distance-based approach they call \textit{line number mapping}~\cite{williams_szz_2008}. The core concept, based on work by Canfora \textit{et al.}~\cite{canfora_identifying_2007}, is that individual lines of code are mapped between revisions by calculating normalized Levenshtein edit distances between all candidate mappings, and considering the pair with the lowest distance to be a valid mapping. Although the mapping is not always correct, the authors claim that the added precision is useful for SZZ. The current implementation of SZZ Unleashed is based on Jaccard distances, but could easily be extended to other measures.

\section{SZZ Unleashed -- Implementation} \label{sec:impl}
SZZ Unleashed is a Java implementation, with some supporting Python scripts, of the SZZ algorithm for git repositories. The implementation is based on the seminal paper by {\'S}liwerski \textit{et al.}~\cite{sliwerski_when_2005} and later enhancements by Williams and Spacco~\cite{williams_szz_2008}. To facilitate interaction with git repositories, SZZ Unleashed uses the JGit library~\cite{sohn_eclipse_2017} maintained by the Eclipse Foundation. Using this library, we reduced the use of text parsing and could work directly on the git revision structure. Working with SZZ Unleashed means following the general SZZ workflow: 

\begin{enumerate}
    \item Extract closed BRs from an issue tracker (prerequisite step)
    \item Link individual BRs to bug-fixing commits (SZZ Phase~1)
    \item Identify bug-introducing commits for the bug-fixing commits (SZZ Phase~2)
\end{enumerate}


We explain the implementation of SZZ Unleashed through a running example on the core repository of the Jenkins project. Jenkins constitutes an appropriate example with numerous contributors, including developers in proprietary organizations, consisting of roughly 1 MLoC (predominantly Java) with a well-managed Jira server for issue tracking. The Jenkins project uses a convention to explicitly state unique identifiers (ID) of BRs in bug-fixing commit messages, i.e., SZZ Phase 1 is straight-forward. Furthermore, the Jira REST API and associated Jira Query Language (JQL) greatly simplifies extraction of BRs.

Detailed instructions to get started with SZZ Unleashed is available in the README on GitHub~\cite{svensson_szz_2018}. Prerequisites to build SZZ Unleashed include Gradle and Java 8. Furthermore, to replicate the running example, Python is needed to run the scripts for extracting defect reports from Jenkin's issue tracker and to process them into the input format used by SZZ Unleashed. Alternatively, users can download and run the Docker image available on GitHub.

\subsection{Phase 1 -- Bug-fixing commits}\label{sec:szzphase1}
As shown in Figure~\ref{fig:szz}, Phase 1 results in a set of bug-fixing commits. First, we need to extract BRs from an issue tracker. For the Jira server used by the Jenkins community, we execute the following JQL query:

\begin{verbatim}
project = JENKINS 
    AND issuetype = Bug 
    AND status in (Resolved, Closed) 
    AND resolution = Fixed 
    AND component = core 
    AND created <= "2018-02-20 10:34"
    ORDER BY created DESC
\end{verbatim}
where $issuetype$ eliminates other types of issues such as feature requests, $status$ eliminates issues that are still open, $resolution$ eliminates duplicated BRs, $component$ excludes issues concerning other repositories, $created$ can is used to set a time interval, and ORDER BY sorts the BRs in reverse chronological order.

The unique IDs of the BRs are used to find bug-fixing commits by executing regular expression (regex) patterns on the git log of the software repository under study. For the Jenkins repository, three different formats for referencing BRs exist, namely JENKINS-XXX, HUDSON-XXX and \#XXX, where XXX is the BR ID. The Python script below specifies the regex pattern that could be used, where $key$ is the ID of a BR formatted as JENKINS-XXX and $nbr$ is the associated number XXX. If there is a match for the \#XXX pattern, we perform an extra regex search to verify that the commit message also contains the word `fix', otherwise the corresponding commit is not considered as bug-fixing.

\begin{verbatim}
String pattern = key + '\D|' + '#' + nbr + \
                '\D|HUDSON-' + nbr + '\D'
\end{verbatim}

The above regex pattern can match multiple commit messages for each BR. SZZ~Unleashed uses another regex to identify the true bug-fixing commit among these matches, i.e., we exclude `merge', `cherry pick', and `nothing' commits. Among the remaining commits, SZZ~Unleashed considers the most recent commit as bug-fixing. The following Python code shows the implementation:

\begin{verbatim}
def commit_selector_heuristic(commits):
    for commit in commits:
        if(re.search('[Mm]erge|[Cc]herry|[Nn]oting', commit)):
            continue
        return commit
    return commits[0]
\end{verbatim}

\subsection{Phase 2 -- Bug-Introducing Commits} \label{sec:szzphase2}
\begin{figure*}
\centering
\includegraphics[width=1.25\columnwidth]{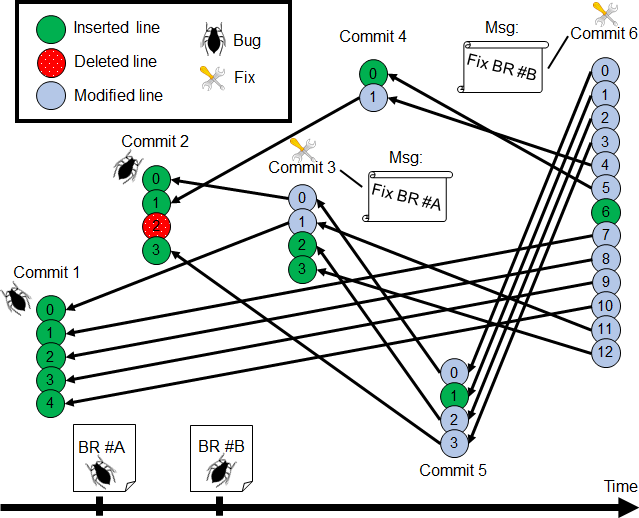}
\caption{Lines of modified code in six commits to an example file. Arrows show the result of line mapping between commits.}
\label{fig:advanced}
\end{figure*}

SZZ Unleashed largely implements the approach Williams and Spacco~\cite{williams_szz_2008}, i.e., line number mappings are used to backtrack through the change history. However, our implementation is not language-specific, thus we do not filter out cosmetic changes. Figure~\ref{fig:advanced} shows an example of SZZ~Unleashed Phase~2. Note that the graph does not show all lines of code, but rather the lines that were altered by the commits. Each changed line of source code is tracked to its creation or a more recent version. How deeply SZZ~Unleashed should traverse the graph is configurable, but the default value is 3. Running git blame on Commit 6 in Figure \ref{fig:advanced} with $depth=1$ would find all but one commit (Commit 2). Using a higher depth setting, however, we can trace to the original commit of any line of code. Thus, with $depth>=3$, node 0 in Commit 6 would be traced to node 0 in Commit 2.

Suppose that Phase 1 of SZZ~Unleashed identified Commit 3 as a bug-fixing commit for BR A. In phase two, git blame is used to identify Commit 2 and Commit 1 -- these would be bug-introducing commit candidates. Next, suppose that Commit 2 was made after BR A was submitted, i.e., it is `Newer' as shown in (D) in Figure~\ref{fig:szz}, and its commit message does not identify it as a partial bug-fix, i.e., there is no explicit reference to a BR. However, Commit 2 can still be bug-introducing for another BR if any of Commit 4, Commit 5 or Commit 6 are bug-fixing commits. Suppose that Commit 6 is a bug-fixing commit for BR B, then Commit 2 will be categorized as a bug-introducing commit -- since it made changes to the same lines of code.

\section{Illustrative Study -- JIT Bug Prediction} \label{sec:example}
This section presents an example of how the output from SZZ~Unleashed can be used. We use the output to train a classifier to identify bug-introducing commits, i.e., JIT bug prediction (referred to as JIT quality assurance by Kemal \textit{et al.}~\cite{kamei_large-scale_2013}). The overall idea is to indicate commits that might require particularly careful code reviews, i.e., providing risk profiles for individual commits.

\subsection{Research goal and method}
How to sample training and test data when evaluating classifiers intended for deployment in issue trackers is critical. First, previous work on supervised learning for issue trackers revealed that disregarding the time dimension, as is done in traditional cross-validation, might lead to overly positive results~\cite{tan_online_2015,jonsson_automated_2016} -- training a classifier on data ``from the future'' is apparently questionable. Second, class imbalance problems might require techniques for oversampling and undersampling, e.g., only 3.6\% of the commits are bug-introducing in our Jenkins dataset (cf. Table~\ref{tab:descStats}). 

We train random forest classifiers to predict bug-introducing commits, i.e, JIT bug prediction. We choose random forest for two reasons: 1) the trained models are reasonably interpretable and 2) Yang \textit{et al.} previously obtained good results in a similar context~\cite{yang_tlel:_2017}. Based on intial trial runs, we set the number of trees to 200. Based on the Jenkins datatset created using SZZ Unleashed, we investigate two research questions:

\begin{itemize}
    \item[RQ1] How does oversampling and undersampling affect the JIT bug prediction for highly imbalanced classes?
    \item[RQ2] Does cross-validation generate better results than a time-sensitive evaluation setup?
\end{itemize}

We investigate RQ1 by comparing a baseline without particular sampling techniques to three approaches that result in an equal proportion of positive and negative training examples -- all available in the sci-kit learn library \textit{imbalanced-learn}~\cite{lemaitre_imbalanced-learn:_2017}. SMOTE oversamples, Cluster Centroids undersamples, and SMOTE+Tomek combines oversampling and undersampling. As shown in Table~\ref{tab:descStats}, our Jenkins dataset contains only 3.6\% positive examples, i.e., the class imbalance problem is more evident than in previous work. 

We study RQ2 by comparing stratified 10-fold cross-validation to ``Online Change Classification'' as described by Tan \textit{et al.}~\cite{tan_online_2015}, i.e., an approach to respect the time dimension when defining training and test data. Using the terminology introduced by the authors, we used the following configuration of time `gaps': SGAP=331, GAP=73, EGAP=781, Update=200, Training duration=1,700, and Test duration=400 (all units in days).

\subsection{Data collection and feature selection}
In line with the description in Section~\ref{sec:impl}, we use SZZ~Unleashed to extract bug-introducing commits from 12 years of development history in the Jenkins core repository (from Nov 5, 2006 until Feb 20, 2018). Table~\ref{tab:descStats} shows descriptive statistics of the resulting dataset and analogous statistics from five datasets collected by Kamei \textit{et al.}~\cite{kamei_large-scale_2013}. Bug-introducing commits and bug-fixing commits are listed as `Bugs' and `Fixes', respectively -- along with commits that are categorized as both. Percentages show the fraction of `Bugs' and `Fixes' among the total number of commits.

\begin{table}
\caption{Descriptive statistics of the extracted Jenkins dataset and five analogous datasets from previous work~\cite{kamei_large-scale_2013}.}
\footnotesize
\centering
\begin{tabular}{|l|c|c|c|c|}
\hline
\textbf{Dataset} & \textbf{\#Bugs} & \textbf{\#Fixes} & \textbf{\#(Fixes $\cap$ Bugs)} & \textbf{\#Commits} \\\hline
\textbf{Jenkins} & \textbf{954 (3.6\%)} & \textbf{2,979 (11.3\%)} & \textbf{808 (3.1\%)} & \textbf{26,378}\\\hline
Bugzilla & 1,696 (36.1\%) & 3,973 (86.0\%) & 1,586 (34.3\%) & 4,620\\\hline
Columba & 1,361 (30.5\%) & 1,463 (32.8\%) & 439 (9.6\%) & 4,455 \\\hline
JDT & 5,089 (14.4\%) & 10,799 (30.5\%) & 2,218 (6.3\%) & 35,386 \\\hline
Mozilla & 5,149 (5.2\%) & 62,888 (64.0\%) & 3,943 (4.0\%) & 98,275 \\\hline
Postgres & 5,119 (25.1\%) & 8,933 (43.7\%) & 2,043 (10.0\%) & 20,431 \\
\hline
\end{tabular}
\label{tab:descStats}
\end{table}

Based on previous work on bug prediction, we represent commits by 16 features as presented in Table~\ref{tab:features}. Ft1--Ft3 are related to code churn as defined by Nagappan and Ball \textit{et al.}~\cite{nagappan_use_2005}, Ft4--Ft13 were all used by Kamei \textit{et al.}~\cite{kamei_large-scale_2013}, and Ft14--Ft16 consider coupling as proposed by D'Ambros \textit{et al.}~\cite{dambros_relationship_2009}. We used code-maat version 1.1 to calculate the values for the coupling features~\cite{tornhill_code-maat_2017}. Table~\ref{tab:features} also shows the relative significance of the 16 features in the random forest classifiers (described next). We observe that the ranking of features is similar to findings from previous work~\cite{moser_comparative_2008,hall_systematic_2012}.

\begin{table}[!ht]
\caption{Features used to represent commits.}
\footnotesize
\centering
\begin{tabular}{|c|l|c|}
\hline
\textbf{ID} & \textbf{Feature} & \textbf{Rel. Sign.}\\\hline
Ft1 & Lines of code added / Total lines of code & 0.17 \\\hline
Ft2 & Lines of code deleted / Total lines of code & 0.04 \\ \hline
Ft3 & Files churned / Number of files & 0.08 \\ \hline
Ft4 & Lines of code in previous version & 0.07 \\\hline
Ft5 & Number of modified subsystems & 0.11 \\\hline
Ft6 & Number of modified sub-directories & 0.09 \\\hline
Ft7 & Entropy (spreading of changes) & 0.16 \\\hline
Ft8 & Purpose of a change (e.g., bug fix) & 0.03 \\\hline
Ft9 & Number of previous committers & 0.08 \\ \hline
Ft10 & Time between committer's contributions & 0.04 \\\hline
Ft11 & Number of unique changes & 0.04 \\\hline
Ft12 & Overall experience of committer & 0.04 \\\hline
Ft13 & Recent experience of committer & 0.03 \\\hline
Ft14 & Number of highly coupled files & 0.00 \\ \hline
Ft15 & Number of coupled files for all degrees & 0.01 \\ \hline
Ft16 & Number of non-modified coupled files & 0.01 \\ \hline
\end{tabular}
\label{tab:features}
\end{table}

\subsection{Results and discussion}
Table~\ref{tab:results} shows the classification accuracy of the random forest classifiers for eight different experimental runs. The table shows precision, recall, and F1 score for two evaluation setups: 1) stratified 10-fold cross-validation and 2) online change classification. For both setups, we report results from applying four different sampling techniques. All values are reported with standard deviations.

\begin{table}
\caption{Classification accuracy for JIT bug prediction.}
\footnotesize
\centering
\begin{tabular}{|l|c|c|c|}
    \hline
	\multicolumn{4}{|c|}{\textbf{Stratified 10-fold Cross-Validation}} \\
    \hline
    Sampling technique & Precision & Recall & F1 score \\
    \hline
    Baseline & 0.156$\pm$ 0.246 & 0.026$\pm$ 0.042 & 0.029$\pm$ 0.034 \\
    \hline
    SMOTE & 0.123$\pm$ 0.076 & 0.212$\pm$ 0.136 & 0.154$\pm$ 0.096 \\
	\hline
    SMOTE+Tomek & 0.117$\pm$ 0.071 & 0.206$\pm$ 0.130 & 0.148$\pm$ 0.091 \\
    \hline
    Cluster Centroids & 0.037$\pm$ 0.001 & 0.945$\pm$ 0.037 & 0.072$\pm$ 0.002 \\
    \hline\hline
    \multicolumn{4}{|c|}{\textbf{Online Change Classification}} \\
    \hline
    Sampling technique & Precision & Recall & F1 score \\
    \hline
    Baseline & 0.210$\pm$ 0.177 & 0.017$\pm$ 0.014 & 0.031$\pm$ 0.026 \\
    \hline
    SMOTE & 0.147$\pm$ 0.041 & 0.104$\pm$ 0.034 & 0.116$\pm$ 0.031 \\
    \hline
    SMOTE+Tomek & 0.163$\pm$ 0.018 & 0.126$\pm$ 0.043 & 0.137$\pm$ 0.030 \\
    \hline
    Cluster Centroids & 0.028$\pm$ 0.004 & 0.917$\pm$ 0.037 & 0.054$\pm$ 0.008 \\
    \hline
\end{tabular}
\label{tab:results}
\end{table}

We find that oversampling is essential for JIT bug prediction subject to the class imbalance problem (RQ1). Both for cross-validation and online change classification, using SMOTE or SMOTE+Tomek obtains considerably higher F1 scores compared to the baseline. Oversampling results in decreased precision, but also substantial improvements in recall. It is clear that using the baseline sampling leads to a too conservative classifier (recall $<3\%$) for the highly imbalanced Jenkins dataset. Note that also undersampling using Cluster Centroids improves recall and F1 score, but the resulting precision ($<4\%$) would never be useful in practice -- probably the resulting training set contains too few examples for the classifier to learn from.

Our investigation of time-sensitivity largely confirms findings from previous work, i.e., disregarding the time dimension in issue trackers might lead to overly positive classifier evaluations (RQ2). Table~\ref{tab:results} shows that F1 scores for cross-validation are higher than for online change classification. Jonsson \textit{et al.} reported that ``\textit{cross-validation consistently yielded higher prediction accuracy than conducting more realistic evaluations on bug reports sorted by the submission date}''~\cite{jonsson_automated_2016} in a study on multi-class classification for bug assignment. Analogous to our work, Tan \textit{et al.} performed binary classification for JIT bug prediction. They concluded that ``\textit{cross-validation presents a false impression of higher precisions}''~\cite{tan_online_2015}. 

On the other hand, our findings partly contrast previous conclusions. Tan \textit{et al.} specifically points out that cross-validations results in falsely high precision results. In our study, however, we instead observe this phenomenon for recall. For sampling using SMOTE and SMOTE+Tomek, cross-validation obtains roughly twice as high recall as the time-sensitive setup. Thus, we conclude that using cross-validation can lead to overly positive results both for precision and recall -- both should be carefully investigated in empirical studies.

On a final note, we do not think JIT bug prediction corresponding to an F1 score of 0.10--0.15 is sufficiently accurate to be of practical value for developers. The false positives would be too many for developers to trust the predictions, and at the same time the classifier would miss too many truly bug-introducing commits. Moreover, the SZZ algorithm has limitations~\cite{rodriguez-perez_reproducibility_2018} -- possibly the classification accuracy could reach the utility break-point if instead a manually annotated training set of commits was used. Nonetheless, such investigations, and a deeper analysis of threats to validity of our small empirical inquiry, is beyond the scope of the illustrative example presented in this section.

\section{Conclusion} \label{sec:conc}
Numerous software engineering studies rely on the SZZ algorithm. Unfortunately, due to the lack of publicly available tool solutions, most researchers must implement their own versions. While the learning process for the individual researcher might be valuable, we argue that the lack of a public SZZ tool might lead to 1) the community reinventing the wheel, 2) hampered reproducability, and 3) research results based on non-disclosed SZZ implementations that might contain bugs.

We respond to the call by Rodriguez \textit{et al.}~\cite{rodriguez-perez_reproducibility_2018} and present \textit{SZZ~Unleashed}, an implementation of the SZZ algorithm publicly available on GitHub under an MIT license. SZZ Unleashed is implemented in Java, with some supporting Python scripts, and includes line number mappings -- an improvement proposed by Williams and Spacco~\cite{williams_szz_2008}. We have already approved the first external pull request and we warmly welcome further contributions from the community.

To illustrate how SZZ Unleashed can be used, both this paper and the GitHub repository are accompanied by an example study of JIT bug prediction using a random forest classifier for the Jenkins project. We report modest classification accuracy (F1 score of roughly 15\%), but corroborate two findings from previous work. First, oversampling is essential in JIT bug prediction for highly imbalanced classes. Second, solely presenting results from cross-validation is not appropriate when evaluating classifiers for software engineering data with timestamps -- there is a high risk of obtaining an excessively positive classification accuracy.

\begin{acks}
Our thanks go to Sven Selberg and Axis Communication AB for hosting the MSc thesis project resulting in this paper. This work has been financially supported by the ITEA3 initiative TESTOMAT Project through Vinnova -- Sweden's innovation agency.
\end{acks}

%% file: EASE19_main.bbl

\begin{thebibliography}{26}


\ifx \showCODEN    \undefined \def \showCODEN     #1{\unskip}     \fi
\ifx \showDOI      \undefined \def \showDOI       #1{#1}\fi
\ifx \showISBNx    \undefined \def \showISBNx     #1{\unskip}     \fi
\ifx \showISBNxiii \undefined \def \showISBNxiii  #1{\unskip}     \fi
\ifx \showISSN     \undefined \def \showISSN      #1{\unskip}     \fi
\ifx \showLCCN     \undefined \def \showLCCN      #1{\unskip}     \fi
\ifx \shownote     \undefined \def \shownote      #1{#1}          \fi
\ifx \showarticletitle \undefined \def \showarticletitle #1{#1}   \fi
\ifx \showURL      \undefined \def \showURL       {\relax}        \fi
\providecommand\bibfield[2]{#2}
\providecommand\bibinfo[2]{#2}
\providecommand\natexlab[1]{#1}
\providecommand\showeprint[2][]{arXiv:#2}

\bibitem[\protect\citeauthoryear{Canfora, Cerulo, and Di~Penta}{Canfora
  et~al\mbox{.}}{2007}]%
        {canfora_identifying_2007}
\bibfield{author}{\bibinfo{person}{G. Canfora}, \bibinfo{person}{L. Cerulo},
  {and} \bibinfo{person}{M. Di~Penta}.} \bibinfo{year}{2007}\natexlab{}.
\newblock \showarticletitle{Identifying {Changed} {Source} {Code} {Lines} from
  {Version} {Repositories}}. In \bibinfo{booktitle}{\emph{Proc. of the 4th
  {International} {Workshop} on {Mining} {Software} {Repositories}}}.
\newblock
\showISBNx{978-0-7695-2950-9}
\urldef\tempurl%
\url{https://doi.org/10.1109/MSR.2007.14}
\showDOI{\tempurl}


\bibitem[\protect\citeauthoryear{Cavalcanti, Silveira~Neto, Machado, Vale,
  Almeida, and Meira}{Cavalcanti et~al\mbox{.}}{2014}]%
        {cavalcanti_challenges_2014}
\bibfield{author}{\bibinfo{person}{Y. Cavalcanti}, \bibinfo{person}{P.
  Silveira~Neto}, \bibinfo{person}{I. Machado}, \bibinfo{person}{T. Vale},
  \bibinfo{person}{E. Almeida}, {and} \bibinfo{person}{S. Meira}.}
  \bibinfo{year}{2014}\natexlab{}.
\newblock \showarticletitle{Challenges and {Opportunities} for {Software}
  {Change} {Request} {Repositories}: {A} {Systematic} {Mapping} {Study}}.
\newblock \bibinfo{journal}{\emph{Journal of Software: Evolution and Process}}
  \bibinfo{volume}{26}, \bibinfo{number}{7} (\bibinfo{year}{2014}),
  \bibinfo{pages}{620--653}.
\newblock
\showISSN{2047-7481}
\urldef\tempurl%
\url{https://doi.org/10.1002/smr.1639}
\showDOI{\tempurl}


\bibitem[\protect\citeauthoryear{Correia}{Correia}{2017}]%
        {correia_old-szz_2017}
\bibfield{author}{\bibinfo{person}{J. Correia}.}
  \bibinfo{year}{2017}\natexlab{}.
\newblock \bibinfo{booktitle}{\emph{old-szz}}.
\newblock
\urldef\tempurl%
\url{https://github.com/intelligentagents/old-szz}
\showURL{%
\tempurl}


\bibitem[\protect\citeauthoryear{Czerwonka, Das, Nagappan, Tarvo, and
  Teterev}{Czerwonka et~al\mbox{.}}{2011}]%
        {czerwonka_crane:_2011}
\bibfield{author}{\bibinfo{person}{J. Czerwonka}, \bibinfo{person}{R. Das},
  \bibinfo{person}{N. Nagappan}, \bibinfo{person}{A. Tarvo}, {and}
  \bibinfo{person}{A. Teterev}.} \bibinfo{year}{2011}\natexlab{}.
\newblock \showarticletitle{{CRANE}: {Failure} {Prediction}, {Change}
  {Analysis} and {Test} {Prioritization} in {Practice} – {Experiences} from
  {Windows}}. In \bibinfo{booktitle}{\emph{Proc. of the 4th {Conference} on
  {Software} {Testing}, {Verification} and {Validation}}}.
  \bibinfo{pages}{357--366}.
\newblock
\urldef\tempurl%
\url{https://doi.org/10.1109/ICST.2011.24}
\showDOI{\tempurl}


\bibitem[\protect\citeauthoryear{D'Ambros, Lanza, and Robbes}{D'Ambros
  et~al\mbox{.}}{2009}]%
        {dambros_relationship_2009}
\bibfield{author}{\bibinfo{person}{M. D'Ambros}, \bibinfo{person}{M. Lanza},
  {and} \bibinfo{person}{R. Robbes}.} \bibinfo{year}{2009}\natexlab{}.
\newblock \showarticletitle{On the {Relationship} {Between} {Change} {Coupling}
  and {Software} {Defects}}. In \bibinfo{booktitle}{\emph{2009 16th {Working}
  {Conference} on {Reverse} {Engineering}}}. \bibinfo{pages}{135--144}.
\newblock
\urldef\tempurl%
\url{https://doi.org/10.1109/WCRE.2009.19}
\showDOI{\tempurl}


\bibitem[\protect\citeauthoryear{de~Freitas~Farias, Novais, Junior,
  da~Silva~Carvalho, Mendonca, and Spinola}{de~Freitas~Farias
  et~al\mbox{.}}{2016}]%
        {de_freitas_farias_systematic_2016}
\bibfield{author}{\bibinfo{person}{M. de Freitas~Farias}, \bibinfo{person}{R.
  Novais}, \bibinfo{person}{M. Junior}, \bibinfo{person}{L. da Silva~Carvalho},
  \bibinfo{person}{M. Mendonca}, {and} \bibinfo{person}{R. Spinola}.}
  \bibinfo{year}{2016}\natexlab{}.
\newblock \showarticletitle{A {Systematic} {Mapping} {Study} on {Mining}
  {Software} {Repositories}}. In \bibinfo{booktitle}{\emph{Proc. of the 31st
  {Annual} {ACM} {Symposium} on {Applied} {Computing}}}.
  \bibinfo{pages}{1472--1479}.
\newblock
\showISBNx{978-1-4503-3739-7}
\urldef\tempurl%
\url{https://doi.org/10.1145/2851613.2851786}
\showDOI{\tempurl}


\bibitem[\protect\citeauthoryear{Engstr\"om, Runeson, and Skoglund}{Engstr\"om
  et~al\mbox{.}}{2010}]%
        {engstrom_systematic_2010}
\bibfield{author}{\bibinfo{person}{E. Engstr\"om}, \bibinfo{person}{P.
  Runeson}, {and} \bibinfo{person}{M. Skoglund}.}
  \bibinfo{year}{2010}\natexlab{}.
\newblock \showarticletitle{A {Systematic} {Review} on {Regression} {Test}
  {Selection} {Techniques}}.
\newblock \bibinfo{journal}{\emph{Information and Software Technology}}
  \bibinfo{volume}{52}, \bibinfo{number}{1} (\bibinfo{year}{2010}),
  \bibinfo{pages}{14--30}.
\newblock
\showISSN{0950-5849}
\urldef\tempurl%
\url{https://doi.org/10.1016/j.infsof.2009.07.001}
\showDOI{\tempurl}


\bibitem[\protect\citeauthoryear{Fenton and Neil}{Fenton and Neil}{1999}]%
        {fenton_critique_1999}
\bibfield{author}{\bibinfo{person}{N. Fenton} {and} \bibinfo{person}{M. Neil}.}
  \bibinfo{year}{1999}\natexlab{}.
\newblock \showarticletitle{A {Critique} of {Software} {Defect} {Prediction}
  {Models}}.
\newblock \bibinfo{journal}{\emph{Transactions on Software Engineering}}
  \bibinfo{volume}{25}, \bibinfo{number}{5} (\bibinfo{year}{1999}),
  \bibinfo{pages}{675--689}.
\newblock
\showISSN{0098-5589}
\urldef\tempurl%
\url{https://doi.org/10.1109/32.815326}
\showDOI{\tempurl}


\bibitem[\protect\citeauthoryear{Godfrey and Zou}{Godfrey and Zou}{2005}]%
        {godfrey_using_2005}
\bibfield{author}{\bibinfo{person}{M. Godfrey} {and} \bibinfo{person}{L. Zou}.}
  \bibinfo{year}{2005}\natexlab{}.
\newblock \showarticletitle{Using {Origin} {Analysis} to {Detect} {Merging} and
  {Splitting} of {Source} {Code} {Entities}}.
\newblock \bibinfo{journal}{\emph{Transactions on Software Engineering}}
  \bibinfo{volume}{31}, \bibinfo{number}{2} (\bibinfo{year}{2005}),
  \bibinfo{pages}{166--181}.
\newblock
\showISSN{0098-5589}
\urldef\tempurl%
\url{https://doi.org/10.1109/TSE.2005.28}
\showDOI{\tempurl}


\bibitem[\protect\citeauthoryear{Hall, Beecham, Bowes, Gray, and Counsell}{Hall
  et~al\mbox{.}}{2012}]%
        {hall_systematic_2012}
\bibfield{author}{\bibinfo{person}{T. Hall}, \bibinfo{person}{S. Beecham},
  \bibinfo{person}{D. Bowes}, \bibinfo{person}{D. Gray}, {and}
  \bibinfo{person}{S. Counsell}.} \bibinfo{year}{2012}\natexlab{}.
\newblock \showarticletitle{A {Systematic} {Literature} {Review} on {Fault}
  {Prediction} {Performance} in {Software} {Engineering}}.
\newblock \bibinfo{journal}{\emph{Transactions on Software Engineering}}
  \bibinfo{volume}{38}, \bibinfo{number}{6} (\bibinfo{year}{2012}),
  \bibinfo{pages}{1276--1304}.
\newblock
\showISSN{0098-5589}
\urldef\tempurl%
\url{https://doi.org/10.1109/TSE.2011.103}
\showDOI{\tempurl}


\bibitem[\protect\citeauthoryear{Jonsson, Borg, Broman, Sandahl, Eldh, and
  Runeson}{Jonsson et~al\mbox{.}}{2016}]%
        {jonsson_automated_2016}
\bibfield{author}{\bibinfo{person}{L. Jonsson}, \bibinfo{person}{M. Borg},
  \bibinfo{person}{D. Broman}, \bibinfo{person}{K. Sandahl},
  \bibinfo{person}{S. Eldh}, {and} \bibinfo{person}{P. Runeson}.}
  \bibinfo{year}{2016}\natexlab{}.
\newblock \showarticletitle{Automated {Bug} {Assignment}: {Ensemble}-based
  {Machine} {Learning} in {Large} {Scale} {Industrial} {Contexts}}.
\newblock \bibinfo{journal}{\emph{Empirical Software Engineering}}
  \bibinfo{volume}{21}, \bibinfo{number}{4} (\bibinfo{year}{2016}),
  \bibinfo{pages}{1533--1578}.
\newblock


\bibitem[\protect\citeauthoryear{Kamei, Shihab, Adams, Hassan, Mockus, Sinha,
  and Ubayashi}{Kamei et~al\mbox{.}}{2013}]%
        {kamei_large-scale_2013}
\bibfield{author}{\bibinfo{person}{Y. Kamei}, \bibinfo{person}{E. Shihab},
  \bibinfo{person}{B. Adams}, \bibinfo{person}{A. Hassan}, \bibinfo{person}{A.
  Mockus}, \bibinfo{person}{A. Sinha}, {and} \bibinfo{person}{N. Ubayashi}.}
  \bibinfo{year}{2013}\natexlab{}.
\newblock \showarticletitle{A {Large}-scale {Empirical} {Study} of
  {Just}-in-{Time} {Quality} {Assurance}}. In
  \bibinfo{booktitle}{\emph{Transactions on {Software} {Engineering}}},
  Vol.~\bibinfo{volume}{39}. \bibinfo{pages}{757--773}.
\newblock
\urldef\tempurl%
\url{https://doi.org/10.1109/TSE.2012.70}
\showDOI{\tempurl}


\bibitem[\protect\citeauthoryear{Kim, Zimmermann, Pan, and James}{Kim
  et~al\mbox{.}}{2006}]%
        {kim_automatic_2006}
\bibfield{author}{\bibinfo{person}{Sunghun Kim}, \bibinfo{person}{Thomas
  Zimmermann}, \bibinfo{person}{Kai Pan}, {and} \bibinfo{person}{Whitehead
  James}.} \bibinfo{year}{2006}\natexlab{}.
\newblock \showarticletitle{Automatic identification of bug-introducing
  changes}. In \bibinfo{booktitle}{\emph{Proc. of the 21st {International}
  {Conference} on {Automated} {Software} {Engineering}}}.
  \bibinfo{pages}{81--90}.
\newblock


\bibitem[\protect\citeauthoryear{Lemaitre, Nogueira, and Aridas}{Lemaitre
  et~al\mbox{.}}{2017}]%
        {lemaitre_imbalanced-learn:_2017}
\bibfield{author}{\bibinfo{person}{G. Lemaitre}, \bibinfo{person}{F. Nogueira},
  {and} \bibinfo{person}{C. Aridas}.} \bibinfo{year}{2017}\natexlab{}.
\newblock \showarticletitle{Imbalanced-learn: {A} {Python} {Toolbox} to
  {Tackle} the {Curse} of {Imbalanced} {Datasets} in {Machine} {Learning}}.
\newblock \bibinfo{journal}{\emph{Journal of Machine Learning Research}}
  \bibinfo{volume}{18}, \bibinfo{number}{17} (\bibinfo{year}{2017}),
  \bibinfo{pages}{1--5}.
\newblock


\bibitem[\protect\citeauthoryear{Moser, Pedrycz, and Succi}{Moser
  et~al\mbox{.}}{2008}]%
        {moser_comparative_2008}
\bibfield{author}{\bibinfo{person}{R. Moser}, \bibinfo{person}{W. Pedrycz},
  {and} \bibinfo{person}{G. Succi}.} \bibinfo{year}{2008}\natexlab{}.
\newblock \showarticletitle{A {Comparative} {Analysis} of the {Efficiency} of
  {Change} {Metrics} and {Static} {Code} {Attributes} for {Defect}
  {Prediction}}. In \bibinfo{booktitle}{\emph{Proc.of the 30th {International}
  {Conference} on {Software} {Engineering}}}. \bibinfo{pages}{181--190}.
\newblock
\showISBNx{978-1-60558-079-1}
\urldef\tempurl%
\url{https://doi.org/10.1145/1368088.1368114}
\showDOI{\tempurl}


\bibitem[\protect\citeauthoryear{Nagappan and Ball}{Nagappan and Ball}{2005}]%
        {nagappan_use_2005}
\bibfield{author}{\bibinfo{person}{N. Nagappan} {and} \bibinfo{person}{T.
  Ball}.} \bibinfo{year}{2005}\natexlab{}.
\newblock \showarticletitle{Use of {Relative} {Code} {Churn} {Measures} to
  {Predict} {System} {Defect} {Density}}. In \bibinfo{booktitle}{\emph{Proc. of
  the 27th {International} {Conference} on {Software} {Engineering}}}.
  \bibinfo{pages}{284--292}.
\newblock


\bibitem[\protect\citeauthoryear{Rahman, Khatri, Barr, and Devanbu}{Rahman
  et~al\mbox{.}}{2014}]%
        {rahman_comparing_2014}
\bibfield{author}{\bibinfo{person}{F. Rahman}, \bibinfo{person}{S. Khatri},
  \bibinfo{person}{E. Barr}, {and} \bibinfo{person}{P. Devanbu}.}
  \bibinfo{year}{2014}\natexlab{}.
\newblock \showarticletitle{Comparing {Static} {Bug} {Finders} and
  {Statistical} {Prediction}}. In \bibinfo{booktitle}{\emph{Proc. of the 36th
  {International} {Conference} on {Software} {Engineering}}}.
  \bibinfo{pages}{424--434}.
\newblock
\showISBNx{978-1-4503-2756-5}
\urldef\tempurl%
\url{https://doi.org/10.1145/2568225.2568269}
\showDOI{\tempurl}


\bibitem[\protect\citeauthoryear{Rodriguez-Perez, Robles, and
  Gonzalez-Barahona}{Rodriguez-Perez et~al\mbox{.}}{2018}]%
        {rodriguez-perez_reproducibility_2018}
\bibfield{author}{\bibinfo{person}{G. Rodriguez-Perez}, \bibinfo{person}{G.
  Robles}, {and} \bibinfo{person}{J. Gonzalez-Barahona}.}
  \bibinfo{year}{2018}\natexlab{}.
\newblock \showarticletitle{Reproducibility and {Credibility} in {Empirical}
  {Software} {Engineering}: {A} {Case} {Study} based on a {Systematic}
  {Literature} {Review} of the use of the {SZZ} algorithm}.
\newblock \bibinfo{journal}{\emph{Information and Software Technology}}
  (\bibinfo{year}{2018}).
\newblock


\bibitem[\protect\citeauthoryear{Rosen, Grawi, and Shihab}{Rosen
  et~al\mbox{.}}{2015}]%
        {rosen_commit_2015}
\bibfield{author}{\bibinfo{person}{C. Rosen}, \bibinfo{person}{B. Grawi}, {and}
  \bibinfo{person}{E. Shihab}.} \bibinfo{year}{2015}\natexlab{}.
\newblock \showarticletitle{Commit {Guru}: {Analytics} and {Risk} {Prediction}
  of {Software} {Commits}}. In \bibinfo{booktitle}{\emph{Proc. of the 2015 10th
  {Joint} {Meeting} on {Foundations} of {Software} {Engineering}}}.
  \bibinfo{pages}{966--969}.
\newblock
\showISBNx{978-1-4503-3675-8}
\urldef\tempurl%
\url{https://doi.org/10.1145/2786805.2803183}
\showDOI{\tempurl}


\bibitem[\protect\citeauthoryear{Sliwerski, Zimmermann, and Zeller}{Sliwerski
  et~al\mbox{.}}{2005}]%
        {sliwerski_when_2005}
\bibfield{author}{\bibinfo{person}{J. Sliwerski}, \bibinfo{person}{T.
  Zimmermann}, {and} \bibinfo{person}{A. Zeller}.}
  \bibinfo{year}{2005}\natexlab{}.
\newblock \showarticletitle{When {Do} {Changes} {Induce} {Fixes}?}. In
  \bibinfo{booktitle}{\emph{Proc. of the 2005 {International} {Workshop} on
  {Mining} {Software} {Repositories}}}, Vol.~\bibinfo{volume}{30}.
  \bibinfo{pages}{1--5}.
\newblock


\bibitem[\protect\citeauthoryear{Sohn, Pearce, Loskutov, Aniszczyk, Halstrick,
  Ranger, Borowitz, Pursehouse, Wagenknecht, Nieder, Sawicki, Kinzler,
  Rosenberg, Stocker, Zivkov, Lay, Parker, and Wolf}{Sohn
  et~al\mbox{.}}{2017}]%
        {sohn_eclipse_2017}
\bibfield{author}{\bibinfo{person}{M. Sohn}, \bibinfo{person}{S. Pearce},
  \bibinfo{person}{A. Loskutov}, \bibinfo{person}{C. Aniszczyk},
  \bibinfo{person}{C. Halstrick}, \bibinfo{person}{C. Ranger},
  \bibinfo{person}{D. Borowitz}, \bibinfo{person}{D. Pursehouse},
  \bibinfo{person}{G. Wagenknecht}, \bibinfo{person}{J. Nieder},
  \bibinfo{person}{K. Sawicki}, \bibinfo{person}{M. Kinzler},
  \bibinfo{person}{R. Rosenberg}, \bibinfo{person}{R. Stocker},
  \bibinfo{person}{S. Zivkov}, \bibinfo{person}{S. Lay}, \bibinfo{person}{T.
  Parker}, {and} \bibinfo{person}{T. Wolf}.} \bibinfo{year}{2017}\natexlab{}.
\newblock \bibinfo{booktitle}{\emph{Eclipse {JGit}}}.
\newblock
\urldef\tempurl%
\url{https://github.com/eclipse/jgit}
\showURL{%
\tempurl}


\bibitem[\protect\citeauthoryear{Svensson and Berg}{Svensson and Berg}{2018}]%
        {svensson_szz_2018}
\bibfield{author}{\bibinfo{person}{O. Svensson} {and} \bibinfo{person}{K.
  Berg}.} \bibinfo{year}{2018}\natexlab{}.
\newblock \bibinfo{booktitle}{\emph{{SZZ} {Unleashed}}}.
\newblock
\urldef\tempurl%
\url{https://github.com/wogscpar/SZZUnleashed}
\showURL{%
\tempurl}


\bibitem[\protect\citeauthoryear{Tan, Tan, Dara, and Mayeux}{Tan
  et~al\mbox{.}}{2015}]%
        {tan_online_2015}
\bibfield{author}{\bibinfo{person}{M. Tan}, \bibinfo{person}{L. Tan},
  \bibinfo{person}{S. Dara}, {and} \bibinfo{person}{C. Mayeux}.}
  \bibinfo{year}{2015}\natexlab{}.
\newblock \showarticletitle{Online {Defect} {Prediction} for {Imbalanced}
  {Data}}. In \bibinfo{booktitle}{\emph{Proc. of the 37th {International}
  {Conference} on {Software} {Engineering}}}. \bibinfo{pages}{99--108}.
\newblock


\bibitem[\protect\citeauthoryear{Tornhill}{Tornhill}{2017}]%
        {tornhill_code-maat_2017}
\bibfield{author}{\bibinfo{person}{A. Tornhill}.}
  \bibinfo{year}{2017}\natexlab{}.
\newblock \bibinfo{booktitle}{\emph{code-maat}}.
\newblock
\urldef\tempurl%
\url{https://github.com/adamtornhill/code-maat}
\showURL{%
\tempurl}


\bibitem[\protect\citeauthoryear{Williams and Spacco}{Williams and
  Spacco}{2008}]%
        {williams_szz_2008}
\bibfield{author}{\bibinfo{person}{C. Williams} {and} \bibinfo{person}{J.
  Spacco}.} \bibinfo{year}{2008}\natexlab{}.
\newblock \showarticletitle{{SZZ} {Revisited}: {Verifying} {When} {Changes}
  {Induce} {Fixes}}. In \bibinfo{booktitle}{\emph{Proc. of the 2008 {Workshop}
  on {Defects} in {Large} {Software} {Systems}}}. \bibinfo{pages}{32--36}.
\newblock


\bibitem[\protect\citeauthoryear{Yang, Lo, Xia, and Sun}{Yang
  et~al\mbox{.}}{2017}]%
        {yang_tlel:_2017}
\bibfield{author}{\bibinfo{person}{X. Yang}, \bibinfo{person}{D. Lo},
  \bibinfo{person}{X. Xia}, {and} \bibinfo{person}{J. Sun}.}
  \bibinfo{year}{2017}\natexlab{}.
\newblock \showarticletitle{{TLEL}: {A} {Two}-layer {Ensemble} {Learning}
  {Approach} for {Just}-in-{Time} {Defect} {Prediction}}.
\newblock \bibinfo{journal}{\emph{Information and Software Technology}}
  \bibinfo{volume}{87} (\bibinfo{year}{2017}), \bibinfo{pages}{206--220}.
\newblock


\end{thebibliography}
